\documentclass[11pt,a4paper]{article}

\usepackage{jheppub}
\usepackage{epsfig,amssymb,amsmath,psfrag}

\usepackage[T1]{fontenc}
\usepackage{psfrag}

\catcode`\@=11
\def \be  {\begin{equation}}
\def \ee  {\end{equation}}
\def \ba  {\begin{eqnarray}}
\def \ea  {\end{eqnarray}}
\def \baa {\begin{eqnarray*}}
\def \eaa {\end{eqnarray*}}
\def \bb  {\begin {thebibliography} }
\def \eb  {\end{thebibliography}}

\def \Im {\mathop{\rm Im}\nolimits}

\def \e  {\mathop{\rm e}\nolimits}
\def\phione{\phi}

\newcommand{\ft}[2]{{\textstyle\frac{#1}{#2}}}

\title{The Excited Hexagon Reloaded}
\subheader{DESY 13-197}

\author[a]{J.~Bartels}
\author[b]{J.~Kotanski}
\author[c]{V.~Schomerus}
\author[c]{M.~Sprenger}

\emailAdd{joachim.bartels@desy.de}
\emailAdd{jan.kotanski@desy.de}
\emailAdd{volker.schomerus@desy.de}
\emailAdd{martin.sprenger@desy.de}

\affiliation[a]{II. Institute for Theoretical Physics,\\ Hamburg University, Germany}
\affiliation[b]{DESY Photon Science,\\ Hamburg, Germany}
\affiliation[c]{DESY Theory Group,\\Hamburg, Germany}

\abstract{This work revisits the computation of six-gluon scattering
amplitudes in the high energy limit of strongly coupled ${\cal N}=4$
supersymmetric Yang-Mills theory. It is based on previous studies in
which we showed that the amplitude simplifies in the Regge regime and
outlined an efficient computational scheme. By exploiting a 
symmetry
of the underlying equations we are now able to argue that a term we
had seen in preliminary numerical studies must vanish identically.
The derived formula for the Regge limit of the 6-gluon scattering
amplitude at strong coupling differs from the one we had conjectured
previously.}


\notoc

\begin{document}
\maketitle
\pagebreak
The computation of all-loop scattering amplitudes in ${\cal N}=4$
supersymmetric Yang-Mills (SYM) theory is an important problem in
quantum field theory. This difficult task could become more
tractable by considering appropriate kinematic limits. One such
limit in which perturbative gauge theory computations are known to
simplify is the multi-Regge or high energy limit. In \cite{Bartels:2010ej}
we observed that the prescription of Alday et al.\ \cite{Alday:2009dv,
Alday:2010vh} to compute scattering amplitudes for $n=6$ external
gluons in strongly coupled ${\cal N}=4$ SYM theory simplifies
drastically at high energies. The analysis was extended
to any number $ n > 6$ gluons in \cite{Bartels:2012gq}. Motivated
by these simplifications in the high energy limit of scattering
amplitudes both at weak and strong coupling we suggested that
they might also occur for intermediate couplings and hence that
the multi-Regge limit could be particularly well suited for an
interpolation between the perturbative gauge theory and the
string theory regimes. In order to constrain any such
interpolation we developed a scheme to compute the high
energy limit of scattering amplitudes in the strongly
coupled gauge theory.

At strong coupling, the leading order of gluon scattering amplitudes may be obtained from
the area of a minimal surface in $AdS_5$ that is approaching the boundary
along a piecewise light-like loop \cite{Alday:2007hr}. As was shown in
\cite{Alday:2009dv,Alday:2010vh}, this geometric construction admits a
beautiful mathematical reformulation in terms of the free energy of a
certain 1-dimensional quantum integrable system. For six external gluons,
the particle densities in the ground state of this 1-dimensional auxiliary
system are determined by solving a system of coupled non-linear integral
equations,
\begin{equation}
  \log Y_a(\theta)=-m_a\cosh\left(\theta-i\phione\right) +  C_a+\sum\limits_{a'}\int\limits_{\mathbb{R}+i\phione}d\theta' K_{aa'}\left(\theta-\theta'\right)\log\left(1+Y_{a'}(\theta')\right),
  \label{eq:Ysys}
\end{equation}
where $a,a'= 1,2,3$ enumerate different species of excitations.
Their masses $m_a$ and chemical potentials $C_a$ are given by
\begin{equation}
m_1 = m = m_3\ \ , \ \ m_2 = \sqrt{2} m \quad , \quad - C_1 = C = C_3 \ \ , \ \
C_2 = 0 \ .
\end{equation}
The kernel functions $K_{aa'}$ encode the interactions through
their relations with the $2 \rightarrow 2$ scattering matrix,
\begin{equation}
(K_{ab}(\theta)) = \frac{1}{2\pi i } \frac{\partial}{\partial \theta}
\left( \begin{array}{ccc}
\log S_1 & \log S_2 & \log S_1\\ \log S_2 & 2 \log S_1 & \log S_2
\\ \log S_1 & \log S_2 & \log S_1
\end{array} \right)  \quad \quad \mbox{with} \quad \quad
\begin{array}{l}
S_1=S_1(\theta) = i \frac{1-ie^\theta}{1+ie^\theta} \\[4mm]
S_2=S_2(\theta) = \frac{2i\sinh\theta-\sqrt{2}}{2i\sinh\theta+\sqrt{2}}
\end{array}
.
\label{eq:kernel}
\end{equation}
The integral equations \eqref{eq:Ysys} also contain a twist
parameter $\phione$ that enters through the argument of the driving
term and through the integration contour. Due to the decay behavior
of the $Y$-functions for large $\mathrm{Re}(\theta)$, the contour
can be shifted as long as no poles are crossed in the process.
The $Y$-functions must satisfy Eqs.\eqref{eq:Ysys} as long
as the argument $\theta$ satisfies $|\Im (\theta - i \phione)|<
\pi/4$. If $\theta$ leaves this strip of width $\pi/2$ around the
integration contour, the value of the  $Y$ function can be
determined through appropriate shift equations.

There is one observation concerning the solutions of Eqs.\eqref{eq:Ysys} that we will play a crucial role later on:
The $Y$-functions enjoy the shifted symmetry
\begin{equation} \label{eq:sym}
Y_a(\theta +i \phione) = Y_a(-\theta + i \phione)\ \ .
\end{equation}
Let us explain this in a bit more detail. Keeping in mind the way we
solve the Y-system numerically, we can prove the symmetry iteratively,
starting with the driving term $Y^{(0)}_a(\theta)=-m_a \cosh(\theta-i
\phione) + C_a$ itself. It is obvious that $Y_a^{(0)}$ obey
Eq.\eqref{eq:sym}. Assume now that the $n$-th iteration
satisfies $Y^{(n)}_a(\theta+i\phione)=Y^{(n)}_a(- \theta+i\phione)$.
We want to show that the same is true for the $(n+1)$-th iteration,
$$
  \log Y^{(n+1)}_{a}(\tilde \theta)=-m_a\cosh\left(\tilde\theta-
  i\phione\right)+  C_a+\sum\limits_{a'}\int\limits_{\mathbb{R}+i\phione}d\theta' K_{aa'}\left(\tilde \theta-\theta'\right)\log\left(1+Y^{(n)}_{a'}(\theta')\right).
  \label{eq:Ysysit}
$$
Setting $\tilde \theta = \theta +i\phione$, the integral contribution
reads
\begin{align} \nonumber
\nonumber  &\int\limits_{\mathbb{R}+i\phione}d\theta' K\left(\theta +i\phione-\theta'\right)\log\left(1+Y^{(n)}(\theta')\right)= \nonumber  \int\limits_{\mathbb{R}}dx K\left(\theta +x\right)\log\left(1+Y^{(n)}(-x+i\phione)\right)=\\
  & = \int\limits_{\mathbb{R}}dx K\left(- \theta-x\right)\log\left(1+Y^{(n)}(x+i\phione)\right)=
  \int\limits_{\mathbb{R}+i\phione}d\theta' K\left(-\theta +i\phione-\theta'\right)\log\left(1+Y^{(n)}(\theta')\right),
  \label{eq:itproof} \nonumber
\end{align}
where we suppressed all indices. In passing from the first to the
second line we have used the symmetry $K(x)=K(-x)$ which is special
to the kernels appearing in the six-point case and we inserted our
assumption on the symmetry for the $n$-th approximation. Since the
final result is the integral contribution for $\tilde \theta=-
\theta +i\phione$, we have demonstrated that the $(n+1)$-th iteration
of all three $Y$-functions has the claimed symmetry.

After these general comments we shall now focus on the Regge regime.
As we have shown in previous work, the high energy limit in the gauge
theory corresponds to sending the mass parameter $m$ to infinity
while keeping $C$ and $\log w := m\sin\phione$ fixed. For large mass $m$,
one can neglect the integral contributions in Eqs.\eqref{eq:Ysys}
unless a solution of $Y_a(\theta_\ast) = -1$ comes close to the
integration contour. Note that the position $\theta_\ast$ depends
on the choice of parameters $m,C,\phione$. Since the system parameters
are related to dual conformal invariant cross ratios $u_a$, the
positions $\theta_\ast$ depend on the kinematics of the process
under consideration. If the Regge limit is performed in the
Euclidean region, i.e.\ $u_3 \rightarrow 1$ while $u_1,u_2
\rightarrow 0$ with $u_1,u_2 >0$, none of the solutions of $
Y_a(\theta_\ast) = -1$ comes close to the real axis so that, in the
large $m$ limit, the $Y$-functions are well approximated by the
driving terms. In order to approach the Regge limit in the
so-called mixed regime where $u_1,u_2 <0$, one has to continue
the cross ratios from the physical regime, which translates into an analytic continuation for the $Y$-system parameters.
During this continuation of the system parameters, it turns out that a few of the solutions $\theta_\ast$
get close to or cross the integration contour. Whenever this happens,
the integral terms can no longer be dropped, i.e.\ they contribute
to the $Y$-functions at the end of the continuation. However, the new terms
are easy to evaluate. In fact, each solution of $Y_a(\theta_\ast)$
that crosses the contour during the continuation contributes terms
of the form $\pm \log S_b(\theta-\theta_\ast)$ to the logarithm of
the $Y$-functions, with the overall sign depending on the direction
in which the contour is crossed. While these additional terms modify
the $Y$-functions, they preserve the symmetry \eqref{eq:sym} as we
have shown above.

In order to reach the mixed region in \cite{Bartels:2010ej} we
continued the cross ratios $u_a$ along the curve
\be
 u_3(\varphi) \ = \ e^{-2i \varphi} u_3,
\label{pathu}
\ee
while keeping both $u_1$ and $u_2$ fixed. Most of the numerical
evaluation was performed for configurations with $u_1=u_2$. These
are associated with the twist parameter $\phione=0$. The dependence
of the system parameters $m = m(\varphi)$ and $C = C(\varphi)$
on the parameter $\varphi$ of the continuation is shown in figures
of our original publication \cite{Bartels:2010ej}.
For these paths, a pair of solutions to $Y_{2\pm 1}(\theta_\ast)=-1$ crosses the
integration contour while a second pair of solutions to $Y_2(\theta_
\ast)=-1$  is seen to collide at the end of the continuation process.
This general pattern does not change when $u_1 \neq u_2$ at least
within the finite region $10^{-2} \leq  u_1/u_2 \leq 10^2$ that we 
scanned numerically.
It should be noted that for $\phione\neq 0$, the mass parameter always 
stays large during the continuation, while $\phione$ always stays small.
$C$ is a finite quantity at the beginning and the end of the continuation, 
but reaches values of $\mathcal{O}(m)$ during the continuation.

Once we accept this general pattern of the migration of solutions,
we can exploit the symmetry \eqref{eq:sym} to obtain precise
analytical expressions for the $Y$-functions $Y'_a$ in the mixed
region. After we have made the mass parameter large to reach the Regge regime,
we can again neglect the integral contributions at the end of the continuation and the equation
governing $Y'_3$ reads
\begin{equation}
  \log Y_3'(\theta)=-m'\cosh\left(\theta-i\phione'\right)+C'+
  \log\left(\frac{S_1(\theta-\theta_{-})}{S_1(\theta-\theta_{+})}\right),
  \label{eq:Y3ac}
\end{equation}
where primes indicate quantities at the end of the continuation
and $\theta_{+}$ is the position of the crossing solution of
$Y_3(\theta)=-1$ that has $\mathrm{Im}(\theta_{+})>0$ (and
analogously for $\theta_{-}$). By definition, we have that
$Y'_3(\theta_+)=-1$ and therefore
\begin{equation}
  i\pi=\log Y_3'(\theta_+)=-m'\cosh(\theta_+-i\phione')+C'+\log\left(\frac{S_1(\theta_+-\theta_{-})}{S_1(0)}\right).
  \label{eq:explY3}
\end{equation}
Since the left-hand side is finite, so is the right-hand side.
As we send $m'$ to infinity, this is only possible if $\theta_+
-\theta_{-}$ approaches a pole of the S-matrix, since $C'$ is a
finite quantity. From the definition of $S_1(x)$,
Eq.(\ref{eq:kernel}), we obtain
\begin{equation}
  \theta_+-\theta_{-}=i\frac{\pi}{2}.
  \label{eq:cons1}
\end{equation}
Taking into account the symmetry \eqref{eq:sym} of the $Y$-functions
we can conclude that
\begin{equation}
  \theta_{-}+\theta_+=2i\phione'.
  \label{eq:cons2}
\end{equation}
Together, Eqs.(\ref{eq:cons1}) and (\ref{eq:cons2}) imply
\begin{equation}
  \theta_{\pm} =\pm i\frac{\pi}{4}+i\phione'\ .
  \label{eq:tptm}
\end{equation}
One may object that the solutions $\theta_{\pm}$ could leave the fundamental
strip $|\mathrm{Im}(\theta-i\phione)| < \pi/4$ and that we have to pick up
contributions of the form $\log\left(1+Y_2\left(\theta\pm i\frac{\pi}{4}
\right)\right)$, which could also compensate the diverging term in Eq.(\ref{eq:explY3}). This, however, is not possible. If we assume, for example,
that $\mathrm{Im}(\theta-i\phione)>\frac{\pi}{4}$, we find that
\begin{align}
  \mathrm{Im}\left(\theta_{-}-i\phione\right)=\mathrm{Im}\left(-\theta_{+}+i\phione\right)=-\mathrm{Im}\left(\theta_{+}-i\phione\right)<-\frac{\pi}{4},
  \label{eq:shiftarg}
\end{align}
and therefore there are no such contributions for $\theta_{-}$.
The analogous statement is true if $\mathrm{Im}(\theta_{-}-i\phione)<-\frac{\pi}{4}$.
We therefore see that we always have at least one position of the solutions $\theta_{\pm}$ for which Eq.(\ref{eq:Y3ac}) holds, which is enough to use our argument.\par
We now turn to $Y_2'(\theta)$, which is governed by the equation
\begin{equation}
  \log Y_2'(\theta)=-\sqrt{2}m'\cosh\left(\theta-i\phione'\right)+\log\left(\frac{S_2(\theta-\theta_{-})}{S_2(\theta-\theta_+)}\right).
  \label{eq:y2ac}
\end{equation}
To find the positions of the solutions to $Y_2'(\theta_\ast)=-1$ at the end
of the continuation we can repeat the above argument to conclude that they
must lie on poles of the function
\begin{equation}
  \frac{S_2\left(\theta+i\frac{\pi}{4}-i\phione'\right)}{S_2\left(\theta-i\frac{\pi}{4}-i\phione'\right)}=\coth\left(\frac{1}{2}(\theta-i\phione')\right)^2.
  \label{eq:poleeq}
\end{equation}
We therefore find that both solutions have to approach $\theta=i\phione'$.
As this coincides with the imaginary part of the integration contour, these two solutions do not give rise to new contributions.
This conclusion deviates from \cite{Bartels:2010ej} where it seemed, for
our numerical investigations performed for $w$ too close to $1$, as if the two solutions would
meet at $\theta_\ast =0$. Now we understand that such a behavior would be
incompatible with the symmetry \eqref{eq:sym}.
New numerical results with larger $\phione$ confirm the above analytical derivation.

From these results we can determine the value of the cross ratios when we
reach the end-point of the continuation. Inserting the expression \eqref{eq:y2ac}
for the $Y$-function $Y_2$ at the end of the continuation we find to leading
order in $\varepsilon'$
$$  u_1 = \frac{Y_2'(-i\frac{\pi}{4})}{1+Y_2'(-i\frac{\pi}{4})}= \gamma \varepsilon' w'
+ \mathcal{O}((\varepsilon')^2) \quad , \quad
    u_2 =  \frac{Y_2'(i\frac{\pi}{4})}{1+Y_2'(i\frac{\pi}{4})}
    = \gamma \frac{\varepsilon'}{w'} + \mathcal{O}((\varepsilon')^2),$$
with $\gamma = - (3 + 2 \sqrt 2)$ and with parameters $w' = \exp(m'\sin\phi')$ and
$\varepsilon' = \exp(-m'\cos\phi')$. Before the continuation, the same
cross ratios take the form $u_1 \sim w \varepsilon$ and $u_2 \sim
w^{-1} \varepsilon$ so that we conclude
\begin{equation} \varepsilon ^\prime\ =\  \gamma^{-1} \varepsilon
  +\mathcal{O}\left(\varepsilon^2\right)\,,\quad w^\prime\ =\
  w+\mathcal{O}\left(\varepsilon^2 \right) \ \ \label{wecprime},
\end{equation}
see \cite{Bartels:2010ej} for more details. With this preparation we
can evaluate the amplitude. The most important contribution comes
from the free energy
\begin{eqnarray*}
 A^{\prime(6)}_{\textrm{free}} & = & \int\limits_{\mathbb{R} + i \phione'}
 \frac{d \theta}{2\pi} m^\prime \cosh (\theta-i\phione') \log \left[ (1+Y^\prime_{1}(\theta)) (1+Y^\prime_{3}(\theta))
(1+Y^\prime_{2}(\theta))^{\sqrt{2}}
\right] \nonumber \\[2mm]
 & & \hspace*{2cm} + m^\prime i \sinh (\theta_+ - i\phione')
 -m^\prime
i \sinh (\theta_- - i\phione') \,.
\label{Afree1}
\end{eqnarray*}
In comparison with the corresponding formula (5.19) in our original
publication \cite{Bartels:2010ej}, we have corrected a sign mistake
and we dropped the last term that seemed to arise from the solutions
of $Y_2(\theta_\ast) = -1$. Its absence is a direct consequence of
the symmetry \eqref{eq:sym}. After these corrections, Eq.(5.20)
of \cite{Bartels:2010ej} takes the form
\begin{eqnarray*}
 A^{\prime(6)}_{\textrm{free} } &=& - \sqrt{2} m^\prime  +
 \mathcal{O}\left(\varepsilon' \right) \ = \ \sqrt{2} \log \varepsilon' +
 \mathcal{O}\left(\varepsilon' \right)\nonumber \\[2mm]
&=&  \sqrt{2}\log\varepsilon - \sqrt{2}\log \gamma +
A^{(6)}_{\textrm{free}} +\mathcal{O}\left(\varepsilon
\right)\,.
\end{eqnarray*}
Following the steps that were performed in \cite{Bartels:2010ej}
we finally arrive at an expression for the 6-gluon remainder
function in the Regge limit of strongly coupled ${\cal N}=4$
SYM theory
 \begin{equation}
\e^{\frac{\sqrt{\lambda}}{2 \pi} R^\prime+i\delta} \ \sim
\ \left(
(1-u_3) \sqrt{\tilde u_1 \tilde u_2}  \right)
^{\frac{\sqrt{\lambda}}{2 \pi} e_2}
\label{leadingstrong}
\end{equation}
with
$$
e_2 \ = \ \left(- \sqrt{2}+ \ft{1}{2} \log (3+2 \sqrt{2}) \right)
\sim -0.533 \,
$$
and where we have factored out a phase factor $e^{i\delta}$ that cancels
the same term with opposite sign in the BDS-part of the amplitude.
Note that the correct value of $e_2$ deviates from the one spelled
out in \cite{Bartels:2010ej} by an internal sign. After correcting this
mistake, the number $e_2$ that appears in the final expression of the
amplitude possesses the same sign as the BFKL eigenvalue $E_2 = -\frac{\lambda}
{2}(2\log 2 -1)$ at weak coupling.

The result Eq.\eqref{leadingstrong} we have obtained in this note deviates
from the expression we had proposed in the original publication in a more
fundamental way: It does not contain the factor depending on the ratio
$u_1/u_2$ that we had proposed previously. This has far reaching
implications. Recall that at weak coupling the correction to the BDS Ansatz for the amplitude can be written as a sum/integral over the quantum numbers
$n$ and $\nu$ of  the 2-dimensional conformal group.
The integral over $\nu$ is dominated by a saddle point at $i\nu =0$.
Assuming that a similar representation of the amplitude also holds at strong coupling, the term depending on $u_1/u_2$ that appeared in \cite{Bartels:2010ej}
suggested that the saddle point moves to infinity at strong coupling.
The absence of this term, however, means that the saddle point appears
at $i\nu =0$, just as it does at weak coupling. This is consistent with
the recent analysis by Simon Caron-Huot in \cite{Caron-Huot:2013fea}.
\medskip

\noindent
{\bf Acknowledgment:} We thank Simon Caron-Huot, Lev Lipatov and especially
Benjamin Basso for stimulating discussions and for challenging the expression
for the hexagon amplitude we had proposed previously, based on insufficient
numerical data. This work was supported by the SFB 676 and by the
People Programme (Marie Curie Actions) of the European Union's Seventh
Framework Programme FP7/2007-2013/ under REA Grant Agreement
No 317089 (GATIS).

\end{document}